\documentclass{jfm}
\usepackage{graphicx}
\usepackage{epstopdf, epsfig}

\newcommand\Reydels{\ensuremath{Re_{\delta_*}}}

\newcommand\Uinfp{U_{\infty}^+}
\newcommand\Reytau{Re_{\tau}}
\newcommand\dd{\mathrm{d}}

\shorttitle{Universal log-law ?}
\shortauthor{P. A. Monkewitz}

\title{Revisiting the quest for a universal log-law and the role of pressure gradient in ``canonical'' wall-bounded turbulent flows}

\author{Peter A. Monkewitz
  \corresp{\email{peter.monkewitz@epfl.ch}}
}

\affiliation{Faculty of Engineering Science, Swiss Federal Institute of Technology (EPFL)
\break CH-1015, Lausanne, Switzerland}

\begin{document}

\maketitle

\begin{abstract}
The trinity of so-called ``canonical'' wall-bounded turbulent flows, comprising the zero pressure gradient turbulent boundary layer, abbreviated ZPG TBL, turbulent pipe flow and channel/duct flows has continued to receive intense attention as new and more reliable experimental data have become available. Nevertheless, the debate on whether the logarithmic part of the mean velocity profile, in particular the K\'arm\'an constant $\kappa$, is identical for these three canonical flows or flow-dependent is still ongoing. In this paper, which expands upon \citet{ICTAM24}, the asymptotic matching requirement of equal $\kappa$ in the log-law and in the expression for the centerline/free-stream velocity is reiterated and shown to preclude a single universal log-law in the three canonical flows or at least make it very unlikely. The current re-analysis of high quality mean velocity profiles in ZPG TBL's, the Princeton ``Superpipe'' and in channels and ducts leads to a coherent description of (almost) all seemingly contradictory data interpretations in terms of TWO logarithmic regions in pipes and channels: A universal interior, near-wall logarithmic region with the same parameters as in the ZPG TBL, in particular $\kappa_{\mathrm{wall}} \cong 0.384$, but only extending from around $150$ to around $10^3$ wall units, and shrinking with increasing pressure gradient, followed by an exterior logarithmic region with a flow specific $\kappa$ matching the logarithmic slope of the respective free-stream or centerline velocity. The log-law parameters of the exterior logarithmic region in channels and pipes are shown to depend monotonically on the pressure gradient.
\end{abstract}


\section{Introduction}
\label{sec:intro}

After the formulation of the mixing length hypothesis by \citet{Prandtl25} and the logarithmic law $U^+ = \kappa^{-1} \ln(y^+) + B$ for a part of the mean velocity in wall-bounded turbulent flows by \citet{vonKarman35} and others, the K\'arm\'an constant $\kappa$ has remained constant for a long time, with a value of $\kappa = 0.41\,$, and has almost acquired the aura of a fundamental physical constant. Here and in the following, ``+''-superscripts denote the usual wall scaling with friction velocity $\widehat{u}_\tau \equiv (\widehat{\tau}_{\mathrm{wall}}/\widehat{\rho})^{1/2}$ and kinematic viscosity $\widehat{\nu}$, where ``hats'' denote dimensional quantities. The main reasons for this long period of constant $\kappa$ has been the limited range of experimental Reynolds numbers and the use of the Clauser chart \citep{Clauser56, Wei05} to determine the wall shear stress in TBL's.

With the advent of wall wires and, more commonly, the oil film technique \citep{TannerBlows76, Fernholzetal96, Segalini13} experimentalists have started to measure $\widehat{u}_\tau$ in TBL's directly and most $\kappa$'s determined from high Reynolds number data have dropped into the 0.38-0.39 range. In pipes and channels, on the other hand, the spread of $\kappa$'s extracted from data has remained rather large, going from the original $\kappa = 0.436\,$ for the Princeton ``Superpipe'' \citep{ZS97} to as low as 0.35-0.36 \citep{NagibChauhan2008} for channels and high aspect ratio ducts. Despite a growing awareness of the effect of the log-law fitting range on $\kappa$ \citep[see e.g.][]{Segalini13} the discrepancies have not gone away and today's turbulence community appears to be divided into essentially two camps: one advocating a universal $\kappa$ such as for instance \citet{MMHS13} and the other advocating flow-specific $\kappa$'s as expressed most clearly by \citet{NagibChauhan2008}.

The problem with a universal $\kappa$ for the log-laws in the ZPG TBL, the channel and the pipe is that the free stream or centerline velocities in all three canonical flows necessarily have to be of the form $\kappa^{-1}\, \ln(\Rey)$ plus a flow dependent constant, with the same unique $\kappa$ (see following sections for the flow specific Reynolds numbers $\Rey$). This is by no means new as it follows directly from Coles' \citep[][]{Coles56} decomposition of the outer velocity into log-law and ``wake'' which can be written as $U^+_{\mathrm{outer}} = \kappa^{-1} \ln(\Rey\, Y)\,+\, W(Y)$, where $Y$ is the outer coordinate defined such that $(\Rey\, Y) = y^+$ and $Y=1$ at the ``edge'' of the TBL or on the centerline. Since there is no natural intermediate length scale between the viscous and outer scales, it follows from basic perturbation theory that, for $Y \to 0$, the outer expansion must asymptotically match the log-law $\kappa^{-1} \ln(y^+)\,+\, W(0)$, while on the centerline $Y=1$ one has $U^+_{\mathrm{CL}} = \kappa^{-1} \ln(\Rey)\,+\, W(1)$ with the \textbf{same} $\mathbf{\kappa}$. This argument for equal log-law and centerline $\kappa$'s carries over to the TBL, but needs to be reformulated when using the outer wall-normal coordinate of \citet{Rotta62}, as in section \ref{sec:zpgtbl} of this paper. The only possible escape from this requirement of equal $\kappa$'s is to make the Coles wake function $\Rey$-dependent, of the form $W(Y, \Rey)$. Such a possibility is explored in appendix \ref{sec:app}, without success.

Since the uncertainty of measured or computed mean centerline and free stream velocities is minimal, the present data reanalysis starts by fitting the available free stream or centerline data $U^+_\infty$ or $U^+_{\mathrm{CL}}$ with $\kappa^{-1} \ln(\Rey) + C$ and then proceeds to the near-wall log-laws. These freestream/centerline $\kappa$'s are clearly different in the  three canonical flows under study. So one must conclude that the log-law parameters, in particular the K\'arm\'an ``coefficients'' $\kappa$, are flow dependent. However, this conclusion is only inevitable if one assumes (without necessity) that $U^+(y^+)$ contains only a single log region. As will be shown in this paper, the overwhelming majority of high quality mean velocity profiles in the Superpipe (section \ref{sec:pipe}) and in channels/ducts (section \ref{sec:channel}) are well described if the log region is split into two distinct parts: A universal interior log region near the wall with the K\'arm\'an constant $\kappa_{\mathrm{int}} \cong 0.384$ of the ZPG TBL, reviewed in section \ref{sec:zpgtbl}. This universal interior log region extends from the buffer layer to $y^+_{\mathrm{int-ext}}$ of around 700 in channels and 500 in pipes, where the pressure gradient starts to contribute to the momentum balance. At $y^+_{\mathrm{int-ext}}$ the channel and pipe velocity profiles switch to flow specific exterior log-laws with flow specific K\'arm\'an ``parameters'' $\kappa_{\mathrm{ext}}$, equal to the respective centerline $\kappa$'s. Only in the ZPG TBL, the interior and exterior log-laws are one and the same.

\section{The mean velocity profile in the ZPG TBL}
\label{sec:zpgtbl}

The point to be reiterated here is the equality in the ZPG TBL of the $\kappa$'s in the log-law and in the expression for the free-stream velocity $\Uinfp(\Reydels)$, with $\Reydels \equiv \widehat{U}_\infty \, \widehat{\delta}_*/\widehat{\nu}$ and $\widehat{\delta}_*$ the displacement thickness.

This is achieved by showing that the mean velocity data set used by \citet{Pitot13}, which is probably the most thoroughly checked data set, together with the data of \citet{KVickyphd} are nearly perfectly matched by a new fit for the total outer velocity $U^+_{\mathrm{outer}}(\eta)$, with $\eta \equiv y^+/\Reydels$ the outer wall-normal coordinate according to \citet{Rotta62}.
\begin{eqnarray}
\label{TBLout}
U^+_{\mathrm{outer}}(\eta) ~ &=& ~ \frac{1}{0.384}\, \ln\Bigg\{ 3.50 \,\Reydels ~ \times \\* &\times& ~ \tanh^{1/2}\left[\left(\frac{5.05\,\eta}{3.50}\right)^2\,\frac{1 + (10.7\,\eta)^4 + (11.9\,\eta)^5 + (10.7\,\eta)^8}{1 + (11.9\,\eta)^5}\right]\Bigg\}  \nonumber \\
\label{TBLoutapprox}
U^+_{\mathrm{outer}}(\eta \ll 1)  &\sim & U^+_{\mathrm{log}} ~ + ~ (11.4\,\eta)^4 + \, \ldots \quad
\mbox{with} \quad U^+_{\mathrm{log}} = \frac{1}{0.384}\, \ln(y^+) + 4.22 \\
\label{TBLinf}
U^+_{\mathrm{outer}}(\eta \to \infty)  &\equiv & U^+_{\infty} = \frac{1}{0.384}\, \ln(\Reydels) + 3.26
\end{eqnarray}

\begin{figure}
\center
\includegraphics[width=0.65\textwidth]{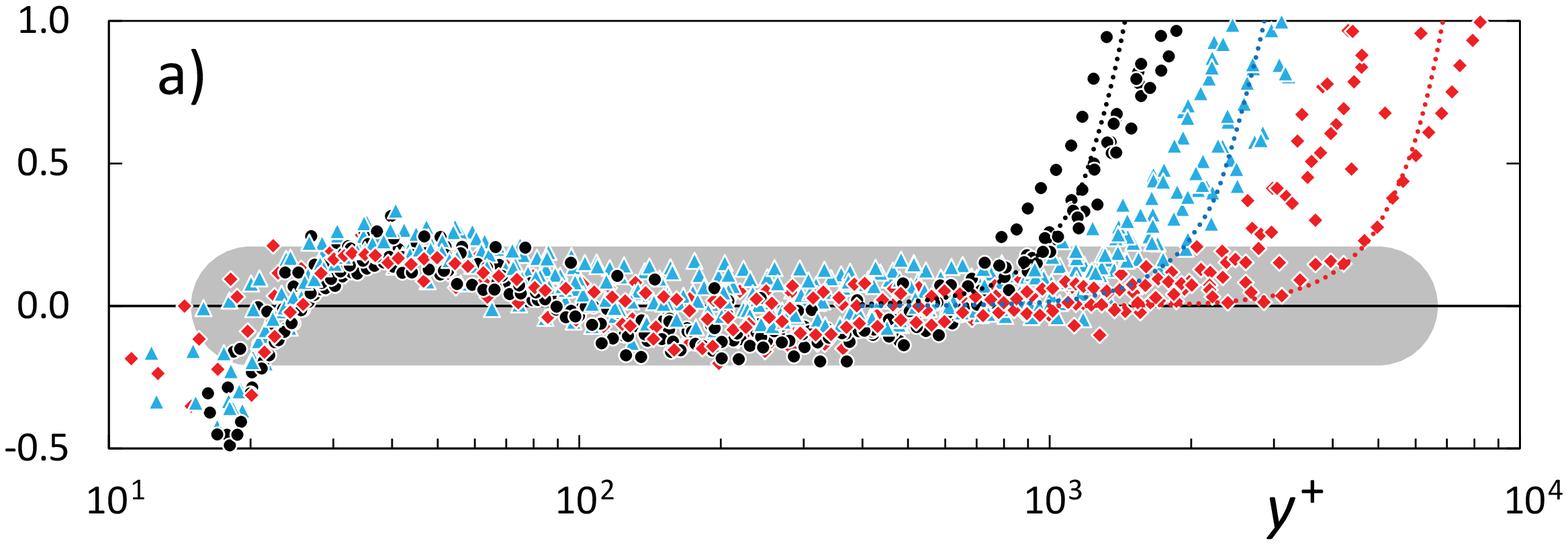}
\includegraphics[width=0.65\textwidth]{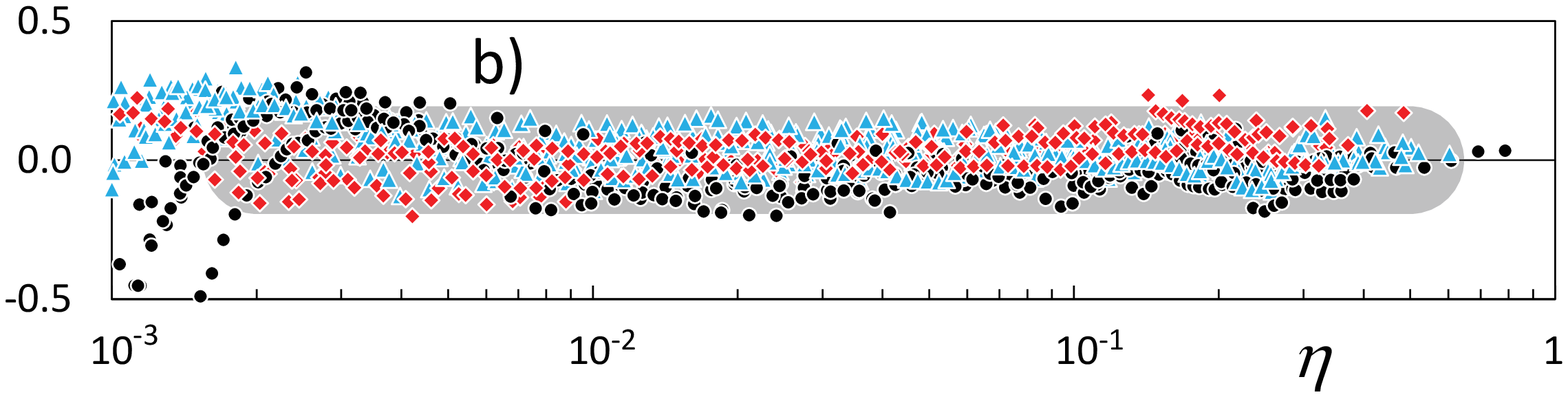}
\caption{(color online) (a) Nineteen ZPG TBL mean velocity profiles used by \citet{Pitot13} to elucidate Pitot corrections and six profiles of \citet{KVickyphd} minus $U^+_{\mathrm{log}}$ (equ. \ref{TBLoutapprox}). $\bullet$ (black), $\Reydels \leq 2\times 10^4$ ; $\blacktriangle$ (blue), $2\times 10^4 < \Reydels \leq 4\times 10^4$ ; $\blacklozenge$ (red), $\Reydels > 4\times 10^4 $. $\cdot\cdot\cdot$, leading term $(11.4\,\eta)^4$ of the small-$\eta$ expansion (\ref{TBLoutapprox}) of $(U^+ - U^+_{\mathrm{log}})$ for $\Reydels = 1.65\times 10^4$ (last black profile), $3.26\times 10^4$ (last blue profile) and $7.84\times 10^4$ (last red profile). (b) Corresponding $(U^+ - U^+_{\mathrm{outer}})$ (equ. \ref{TBLout}) versus $\eta$. The grey bands indicate deviations of up to $\pm 0.2$ from the log-law in (a) and from zero in (b). }
\label{Fig:TBLlog}
\end{figure}

The fit (\ref{TBLout}) is simpler than most previous fits, avoids the introduction of a rather arbitrary boundary layer thickness $\delta$, has equal $\kappa$'s in $U^+_{\mathrm{log}}$ (equ. \ref{TBLoutapprox}) and $U^+_{\infty}$ (equ. \ref{TBLinf}) and satisfies the asymptotic consistency requirement $\int_0^{\infty}\,(\Uinfp - U^+_{\mathrm{outer}})\,d\eta = 1$ within 0.15\% \citep[see][]{MCN07}. The near perfect data fit can be appreciated in figure \ref{Fig:TBLlog}. Note also that the fit (\ref{TBLout}) is likely the first to correctly reproduce the departure from the log-law $\propto \eta^4$ at large $y^+$, shown in fig. \ref{Fig:TBLlog}a and achieved by the ``trick'' of using the square root $\tanh^{1/2}(\eta^2 ...)$ in equation (\ref{TBLout}).

Before moving on to the analysis of pipe and channel data, a few comments are called for. Here and in the following, the value of $\kappa = 0.384$ determined by \citet{MCN07} is used for the ZPG TBL and the interior log-laws to be identified in pipes and channels. It is clear that the third digit of $\kappa$ depends on the choice of data sets, but its value is now generally thought to be in the interval $[0.38, 0.39]$ \citep[see e.g.][]{Metal2010}. As the standard oil film technique to measure wall shear stress leads to a slight systematic over-estimate of $\kappa$ \citep{Segalinietal2015}, the value of $0.384$ is preferred over $0.39$.

\begin{figure}
\center
\includegraphics[width=0.65\textwidth]{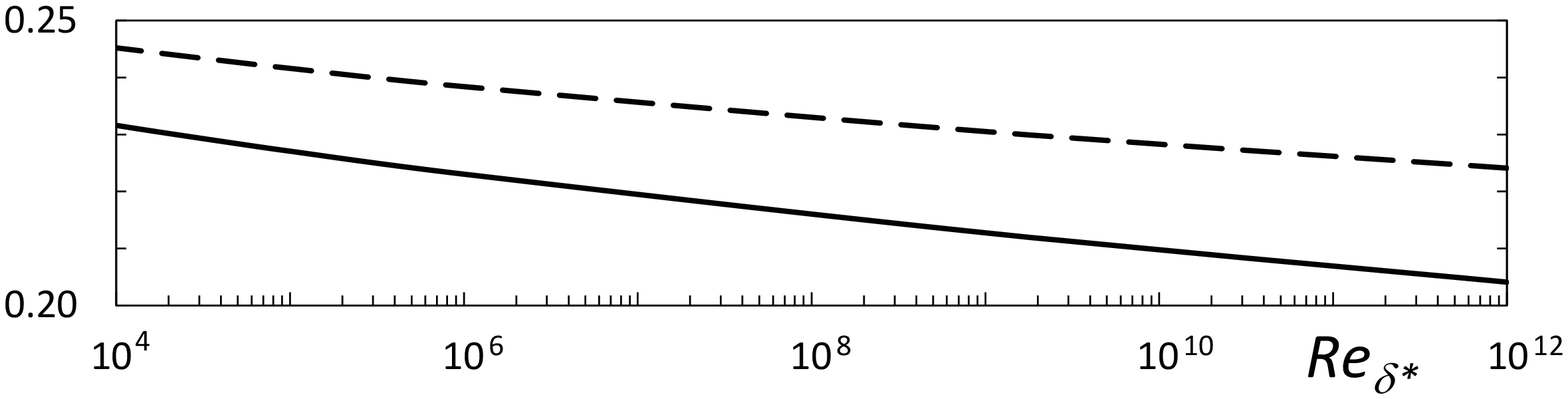}
\caption{ Boundary layer thicknesses $\eta_{.99}$ (---) and $\eta_{.995}$ (- -) versus \Reydels for the outer fit (\ref{TBLout}). }
\label{Fig:d99}
\end{figure}

Finally, since past comparisons with pipe and channel have often been made on the basis of equal $\Reytau$, variously defined in the TBL as $\eta_{.99}\times \Reydels$ or $\eta_{.995}\times \Reydels$, it is useful to recall that these quantities, besides being strongly affected by measurement uncertainties and/or data fits, do not correspond to fixed values of the outer variable $\eta$. In other words, scaling arguments based on $\Reydels$ and $\Reytau$ are not equivalent if they are supposed to be valid to infinite Reynolds number. For the present fit (\ref{TBLout}),  $\eta_{.99}$ and $\eta_{.995}$ decrease significantly from $0.232$ and $0.245$ at $\Reydels = 10^4$ to, respectively, $0.213$ and $0.231$ at $\Reydels = 10^9$, as shown in figure \ref{Fig:d99}.

\section{The mean velocity profile in pipes, with emphasis on the Princeton Superpipe}
\label{sec:pipe}

There exist a large number of turbulent pipe flow experiments, the best known early study being the one by \citet{Nikuradse-pipe}. As the aim of this section is to investigate the relation between the $\kappa$'s extracted from the logarithmic region of $U^+(y^+)$ and from the centerline velocity $U^+_{\mathrm{CL}}(R^+)$, with $R^+ \equiv (\widehat{R} \widehat{u}_{\tau}/\widehat{\nu}) \equiv \Reytau$ the non-dimensional pipe radius and relevant Reynolds number, the requirements on the data are severe. On the one hand the Reynolds numbers must be sufficiently high to obtain a substantial log-law: assuming that the log-law covers the interval $200 \leq y^+ \leq 0.15\,R^+$, a single decade of log-law requires a $R^+ \gtrsim 10^4$. In addition, the data must cover a large range of $R^+$: As $U^+_{\mathrm{CL}}$ is asymptotically $\propto \ln(R^+)$, one decade of $R^+$ corresponds to $U^+_{\mathrm{CL}}$ changing by only about 5, so that an uncertainty of $\pm 0.2$ in $U^+_{\mathrm{CL}}$ translates to a $7\%$ uncertainty in $\kappa$. For these reasons, the following data analysis will focus on the Superpipe data (ZS97, ZS98).

Other investigations of high Reynolds number pipe flow are under way, but have not yet produced data sets comparable to the Superpipe. Among them is an effort by a Japanese group \citep{Furuichi15} which has so far published mean velocity data only up to $R^+ \approx 10^4$ (corresponding to the lowest blue Superpipe profile in figure \ref{Fig:Plog1}). Furthermore, the publication concentrates on friction factor correlations based on bulk velocity and does not directly provide data for the centerline velocity. Another experiment with a Reynolds number range comparable to that of the Superpipe is currently starting up in Italy \citep{CICLOPE} and the first data by \citet{CiclopeRS} hint at a discrepancy between the $\kappa$'s determined from the logarithmic region of $U^+$ and from the centerline velocity $U^+_{\mathrm{CL}}$, an observation already made by \citet{zanoun2007} despite a limited Reynolds number range of $R^+ < 9000$.

The Superpipe data have had a major impact on turbulence research due to the dramatic increase of $R^+$ over previous studies, but two features of these experiments have given rise to much scrutiny and debate :
\begin{itemize}
\item[$\bullet$] One problem was the very large diameter of the Pitot probes in viscous units which decrease to $0.6 \mu$m at the highest $R^+$, for which the schemes to correct for mean shear and turbulence intensity had to be extended into uncharted territory. This issue has been addressed by \citet{mckeonstatic02} and \citet{mckeonpitot03}, using the argument that $U^+(y^+)$ in the wall region is universal and must be independent of the Pitot probe diameter, and more recently by an international collaboration \citep{Pitot13}, which relied on the comparison between Pitot and hot-wire measurements in ZPG TBL's. In addition, \citet{VinDN16} have  investigated the effect of possible Pitot positioning errors. These different efforts have led to a fair agreement on the most appropriate corrections, albeit the uncertainty of the Superpipe data remains relatively large for $y^+ \lesssim 10^3$ in comparison with data sets at lower Reynolds numbers, such as those included in figure \ref{Fig:PCL}. Beyond $y^+ \approx 10^3$, however, the corrections quickly become small.
\item[$\bullet$] The second issue was the influence of roughness at the higher $R^+$ which has been thoroughly studied by \citet{Allen05}. As a result, there is now widespread agreement that only the three highest Reynolds number profiles ($R^+ \gtrsim 2\times 10^5$) are significantly affected by roughness.
\end{itemize}

\subsection{The log-law for the pipe centerline velocity}
\label{sec:pipeCL}

\begin{figure}
\center
\includegraphics[width=0.65\textwidth]{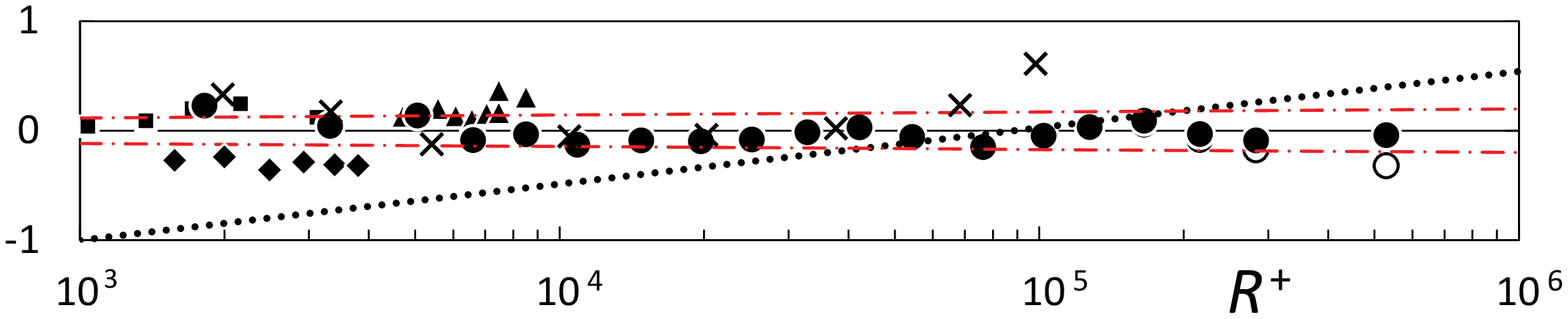}
\caption{(color online) Experimental pipe centerline velocities minus $U^+_{\mathrm{CL}}$ (equ. \ref{PCL}) versus $R^+$. $\bullet$, Superpipe Pitot data \citep{ZS97} with a Hama-like roughness correction $(2 \kappa)^{-1} \ln[1 + (0.14\,k_s^+)^2]$, $\,\widehat{k}_s = 0.45\,\mu$m; $\circ$, same data without roughness correction; $\times$, Superpipe NSTAP data of \citet{Hultetal12}; $\blacklozenge$, \citet{PA77}; $\blacktriangle$, \citet{zanoun2007}; $\blacksquare$, \citet{Monty_thesis}. $\cdot - \cdot$ (red), $\pm 0.5\%$ of $U^+_{\mathrm{CL}}$ (equ. (\ref{PCL}). $\cdot\cdot\cdot$, slope corresponding to $\kappa =0.384$.}
\label{Fig:PCL}
\end{figure}

Starting with the centerline velocity $U^+_{\mathrm{CL}}$ and excluding the lowest as well as the three highest $R^+$, where the roughness correction increases the uncertainty, the Pitot data are least-squares fitted by
\begin{equation}
U^+_{\mathrm{CL}} = \frac{1}{0.42}\, \ln(R^+) + 6.84
\label{PCL}
\end{equation}
with an $R^2$ value of 0.9993. As can be appreciated in figure \ref{Fig:PCL}, all the Superpipe Pitot data beyond the lowest Reynolds number are within $\pm 0.5\%$ of the fit (\ref{PCL}), if corrected for roughness with a Hama-like roughness correction $(2 \kappa)^{-1} \ln[1 + (0.14\,k_s^+)^2]$ and $\,\widehat{k}_s = 0.45\,\mu$m, in line with the investigation of \citet{Allen05}. If all 19 data points are used, the least-squares $\kappa$ increases insignificantly to 0.423.

Comparing with other experiments, it is remarkable that the fit (\ref{PCL}) is essentially identical to the fit of $U^+_{\mathrm{CL}}$ in figure 38 of \citet{Nikuradse-pipe} over the range $10^2 \leq R^+ \leq 5\times 10^4$ and given as $(1/0.417) \ln(R^+) + 6.84$ (note the typo 5.84 instead of 6.84 for the additive constant on p. 66 of the NASA translation). The $\kappa$ in (\ref{PCL}) is also virtually identical to the $\kappa = 0.421$ fitted by \citet{MLJMS04} to the logarithmic region beyond $y^+ \gtrsim 600$ and is consistent with the values estimated by \citet[see in particular their fig. 4b]{Bailey14}, except for the values extracted from the new NSTAP ``micro-hotwire'' data \citep{Hultetal12}. The latter data are also included in figure \ref{Fig:PCL} to support the opinion of this author that the NSTAP technology is not yet sufficiently validated to reliably deduce K\'arm\'an constants : Using all eight NSTAP data points yields a least-squares $\kappa$ of 0.41, while the four lowest and highest $R^+$ yield $\kappa$'s of 0.47 and 0.36, respectively. Finally, it is noted that the various data sets in figure \ref{Fig:PCL} below $R^+$ of $10^4$ are compatible with $\kappa = 0.42$, but too short to extract a reliable value.

\subsection{The ``interior'' and ``exterior'' log-laws}
\label{sec:pipelog}

Turning the attention to the near-wall region of $U^+(y^+)$, the Superpipe data are less helpful below $y^+ \approx 10^3$ for the reasons discussed at the beginning of section \ref{sec:pipe}. At low $y^+$, below around 600, both \citet{ZS98} and \citet{MLJMS04} see a power law, but the latter authors remark that in the interval $350 < y^+ < 950$ a log-law with $\kappa = 0.385$ and an additive constant of 4.15 ``fits the data quite well''.

From the point of view of balance of terms in the Reynolds averaged momentum equation, the mean velocities in pipe and channel flows should, in wall units, be virtually identical to the ZPG TBL mean velocity in the region where the viscous and Reynolds stresses dominate over the pressure gradient. In pipe flow this is the region outside a very thin wall sub-layer in which the viscous stress is balanced by the pressure gradient \citep[see][and references therein]{Klewickietal14}. The thickness $y^+_{\mathrm{pgsl}}$, with ``pgsl'' for pressure gradient sub-layer, can be characterized by the wall distance at which the pressure gradient term is equal to the $(y^+)^4$ term in the Taylor expansion of the ZPG TBL mean velocity originating from the balance with the leading term $\propto (y^+)^3$ of the Reynolds stress. Using table 2 of \citet{MonkNagib2015}, one obtains $y^+_{\mathrm{pgsl}} \approx 10^2/\sqrt{R^+}$ which is so small that this layer has no measurable effect on experimental velocity profiles much beyond $y^+_{\mathrm{pgsl}}$. In the ZPG TBL, on the other hand, the convective terms and the stream-wise stress derivatives are in the inner, near-wall region of order $\mathcal{O}[\Reydels \ln^2(\Reydels)]^{-1}$ \citep[see][]{MonkNagib2015} so that the dominant balance is also between viscous and Reynolds stresses.

Therefore, one expects to see in the pipe (and in the channel) the same log-law beyond $y^+ \approx 150$ with a $\kappa$ close to 0.384 as in the ZPG TBL (see figure \ref{Fig:TBLlog}a), extending over the layer in which the pressure gradient in the momentum equation remains sufficiently smaller than the viscous and Reynolds stresses. Indeed, such a log-law with $\kappa$'s in the range 0.38-0.39 has been clearly identified in many pipe flow studies at lower $R^+ \lesssim 10^4$, where the log-law does not extend beyond $y^+ \approx 10^3$ and probe correction problems are minor compared to the Superpipe : \citet{Monty_thesis}, for instance, identified a log-law for $y^+ \gtrsim 150$ with  $\kappa = 0.384$ and $0.386$, using hot-wires and Pitot tubes, respectively, \citet{zanoun2007} found $\kappa$'s between 0.38 and 0.39 depending on the fitting range, \citet{Furuichi15} deduced a $\kappa$ of $0.382$ from their profiles and \citet{CiclopeRS} found $0.39$ from near-wall profiles for $R^+ \leq 3.2\times 10^4$.

There is no reason for the near-wall $U^+(y^+)$ in the Superpipe to be different from the profiles in experiments with $R^+$ between $10^3$ and $10^4$, such as the ones mentioned above. One must therefore conclude that the Superpipe profiles also have an interior log region with a $\kappa$ of around 0.384 which is just somewhat obscured by the residual uncertainty of Pitot corrections.

\begin{figure}
\center
\includegraphics[width=0.65\textwidth]{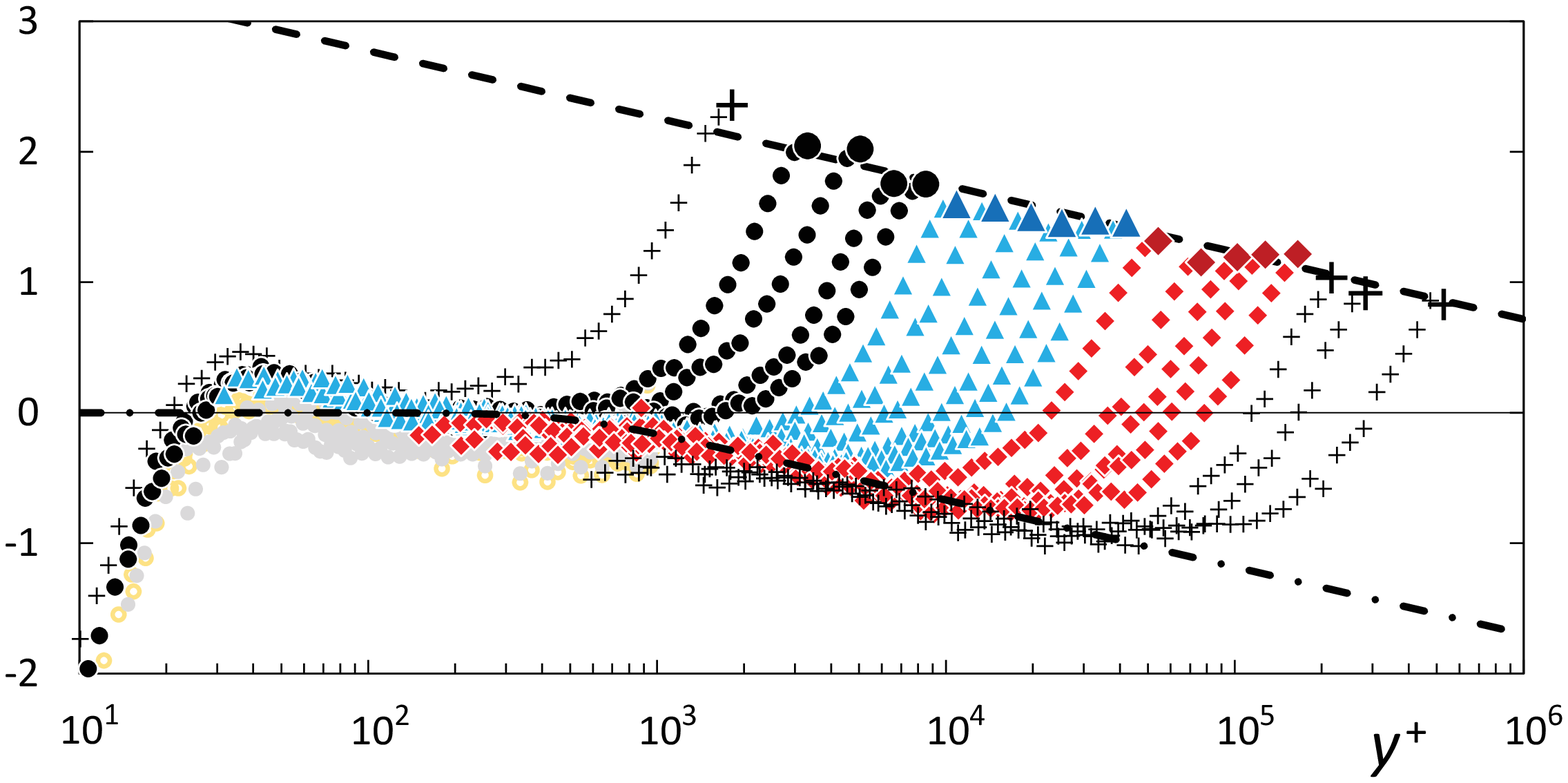}
\caption{(color online) Nineteen Superpipe profiles minus the interior log-law (\ref{Plogint}) $(U^+ - U^+_{\mathrm{logint}})$ versus $y^+$, with Pitot corrections according to McKeon and the Hama-like roughness correction given in the caption of fig. \ref{Fig:PCL}. +, $R^+ = 1.82\times 10^3$ considered low Reynolds number ; $\bullet$ (black), $R^+ = \{3.31, 5.08, 6.58, 8.49\}\times 10^3$ ; $\blacktriangle$ (blue), $R^+ = \{1.09, 1.48, 1.97, 2.51, 3.27, 4.21\}\times 10^4$ ; $\blacklozenge$ (red), $R^+ = \{0.542, 0.761, 1.02, 1.27, 1.65\}\times 10^5$ ; +, $R^+ = \{2.16, 2.83, 5.27\}\times 10^5$ where roughness effects become significant. Corresponding large symbols mark the centerline. $\bullet$ (grey), $\circ$ (yellow), same data for $3\times 10^3 < R^+ < 2\times 10^5$ corrected according to Vinuesa-Nagib and Bailey, respectively. $- -$, $U^+_{\mathrm{CL}} - U^+_{\mathrm{logint}}$ for $R^+ \gg 1$ ; $-\cdot -$, $(-\Delta U^+_{\mathrm{log}})$ given by equ. (\ref{UPlift}) for $R^+ \gg 1$. }
\label{Fig:Plog1}
\end{figure}

To reveal this interior log-law, the ZPG TBL log-law is first modified to satisfy the symmetry condition on the centerline $Y \equiv y^+/R^+ = 1\,$ :
\begin{eqnarray}
\label{Plogint}
U^+_{\mathrm{logint}}(Y, R^+)  &=& \, \frac{1}{0.384}\, \ln \left\{\frac{5.05}{0.5\,\pi}\,R^+\,\sin \left(\frac{\pi\,Y}{2}\right)\right\} \\
&\sim & \,\frac{1}{0.384}\,\ln(y^+)\,+ \,4.22 \, - \, 1.07\, Y^2 \, + \,\mathcal{O}(Y^4) \quad \mbox{for} \quad Y \ll 1 \quad ,\nonumber
\end{eqnarray}
a modification which alters $U^+_{\mathrm{log}}$ in equation (\ref{TBLoutapprox}) only for $Y \gtrsim 0.3$. The result of subtracting $U^+_{\mathrm{logint}}$ (equ. \ref{Plogint}) from the Superpipe Pitot data, corrected according to McKeon \citep[][]{mckeonpitot03}, is shown in figure \ref{Fig:Plog1}. The implementations of the Pitot corrections by Bailey \citep[][]{Pitot13} and Vinuesa-Nagib \citep[][]{Vinuesa2016300}, included in figure \ref{Fig:Plog1} for $y^+ \leq 10^3$, are seen to mainly shift the graphs down by $0.1 - 0.15$.

The most striking feature of figure \ref{Fig:Plog1} is the rather abrupt switchover at $y^+ = \mathcal{O}(10^3)$ of all the profiles with $R^+ \gtrsim 10^4$ to an exterior log-law with the same logarithmic slope $(1/0.420)$ as in equation (\ref{PCL}) for $U^+_{\mathrm{CL}}$. Note that the data for $R^+ < 10^4$  cannot show this second exterior log region, because $y^+ = 10^3$ corresponds already to a $Y \geq 0.1$ which is at or beyond the upper end of any log-law.

The smooth switch from interior to exterior log-law in figure \ref{Fig:Plog1} is well modelled by
\begin{eqnarray}
\label{UPlift}
\Delta U^+_{\mathrm{log}} &=& \frac{1}{3}\left(\frac{1}{0.384} - \frac{1}{0.420}\right) \ln\left\{1 + \left[\frac{0.002}{0.5\,\pi}\,R^+ \sin\left(\frac{\pi Y}{2}\right)\right]^3\right\} \\*
&\sim & 0.223 \,\ln\left\{0.002\,y^+\right\} - 0.0918 \,Y^2  + \mathcal{O}\left(Y^4\right) \quad \mbox{for} \quad \left(0.002 \, R^+\right)^{-1} \ll Y \ll 1 \quad , \nonumber
\end{eqnarray}
which places the boundary between the two log-laws at $y^+ = 1/0.002 = 500$.

\begin{figure}
\center
\includegraphics[width=0.65\textwidth]{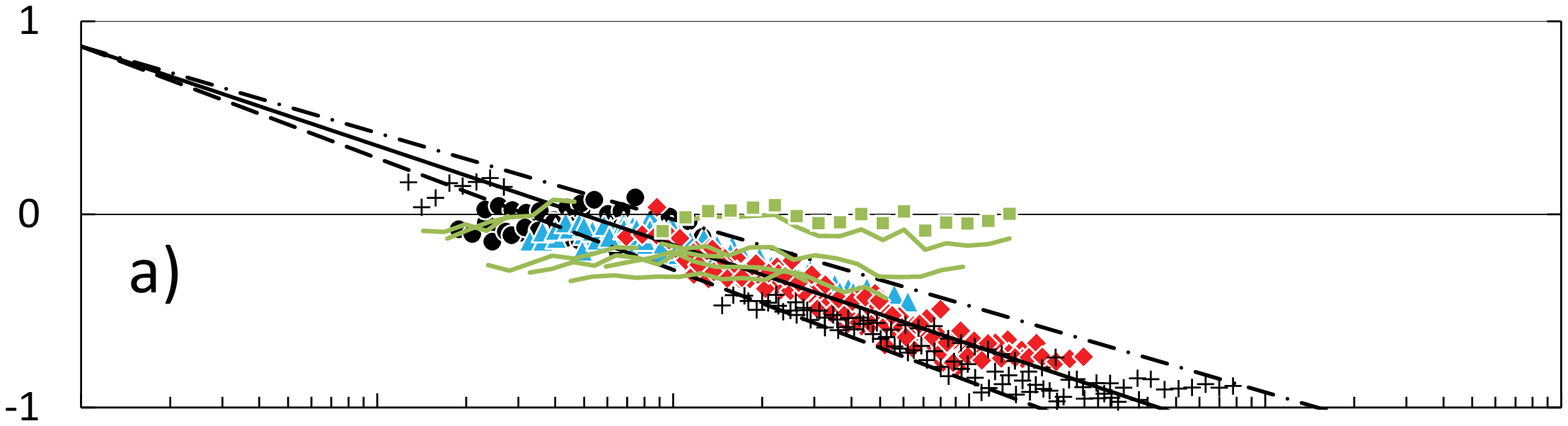}
\includegraphics[width=0.65\textwidth]{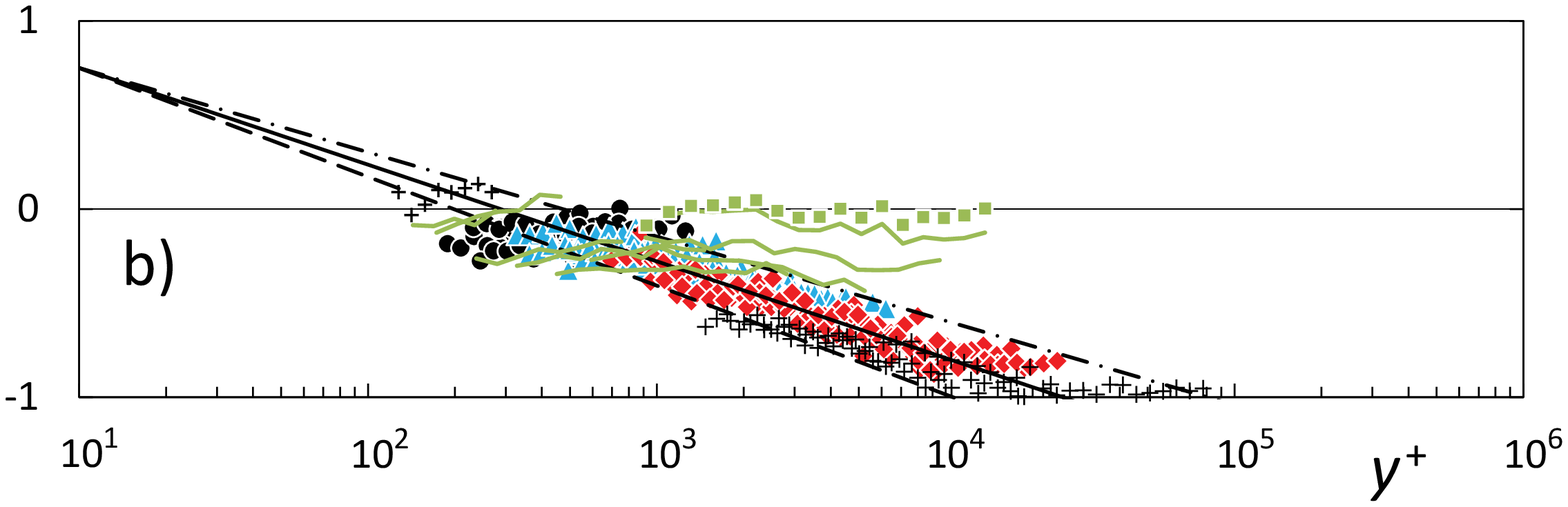}
\caption{(color online) $(U^+ - U^+_{\mathrm{logwall}})$ (equ. \ref{Plogint}) versus $y^+$ for $3(R^+)^{1/2} \leq y^+ \leq 0.15 R^+$. (a) Symbols of fig. \ref{Fig:Plog1}: Pitot data corrected according to \citet{mckeonpitot03}; --- (green), original NSTAP data from \citet{Hultetal12}; $\blacksquare$ (green), NSTAP data for $R^+ = 9.82\times 10^4$ from \citet{MMHS13}. Slopes corresponding to $\kappa = 0.425$ ($- -$), $0.42$ (---) and $0.415$ ($-\cdot -$). (b) Analogous to (a) but with Pitot data corrected according to \citet{Pitot13} .}
\label{Fig:Plog2}
\end{figure}

To reinforce the point about the Pitot corrections becoming non-controversial beyond $y^+ \approx 10^3$, the Pitot data, corrected according to both McKeon and Bailey, minus the interior log-law (\ref{Plogint}) are shown in figure \ref{Fig:Plog2} for the interval $3(R^+)^{1/2} \leq y^+ \leq 0.15 R^+$. This is the extent of the unique and universal logarithmic region with $\kappa = 0.39$, claimed by \citet{MMHS13} to exist in all three canonical flows. To understand this claim, the Superpipe data in their figure 1, together with the original NSTAP data of \citet{Hultetal12}, are included in figure \ref{Fig:Plog2}. These data, obtained with a radically new probe for which experience is still limited, are seen to be the only ones supporting such a universal log-law with $\kappa = 0.39$ beyond $y^+$ of $10^3$. Since a $U^+_{\mathrm{CL}}$ of the form (\ref{PCL}) with $\kappa = 0.39$ is not defendable on the basis of figure \ref{Fig:PCL}, as discussed in section \ref{sec:pipeCL}, the universal single log-law proposed by \citet{MMHS13} could only be ``saved'' by a Reynolds-dependent correction to the usual wake function. A failed attempt to find such an alternative wake function is reported in appendix \ref{sec:app}. Although this failure does not represent any proof, it strongly supports the present description in terms of an interior log-law with a universal $\kappa$ close to 0.384, common to all wall-bounded flows with weak pressure gradients, and, at high enough Reynolds numbers, an exterior log-law with pressure-gradient dependent $\kappa$ (see the discussion in section \ref{sec:conclusion}).

\subsection{Fitting the wake region}
\label{sec:pipewake}

\begin{figure}
\center
\includegraphics[width=0.65\textwidth]{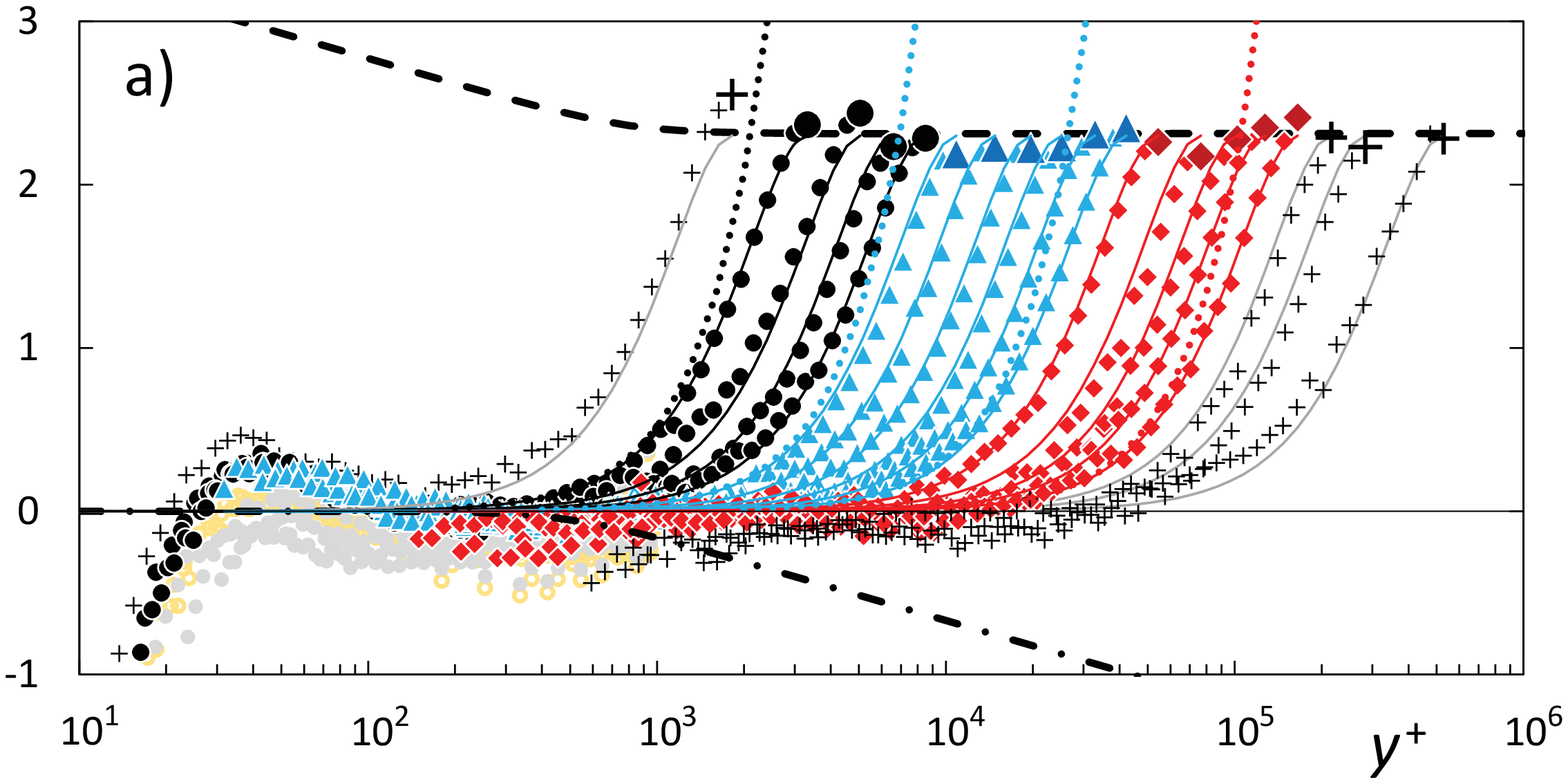}
\includegraphics[width=0.65\textwidth]{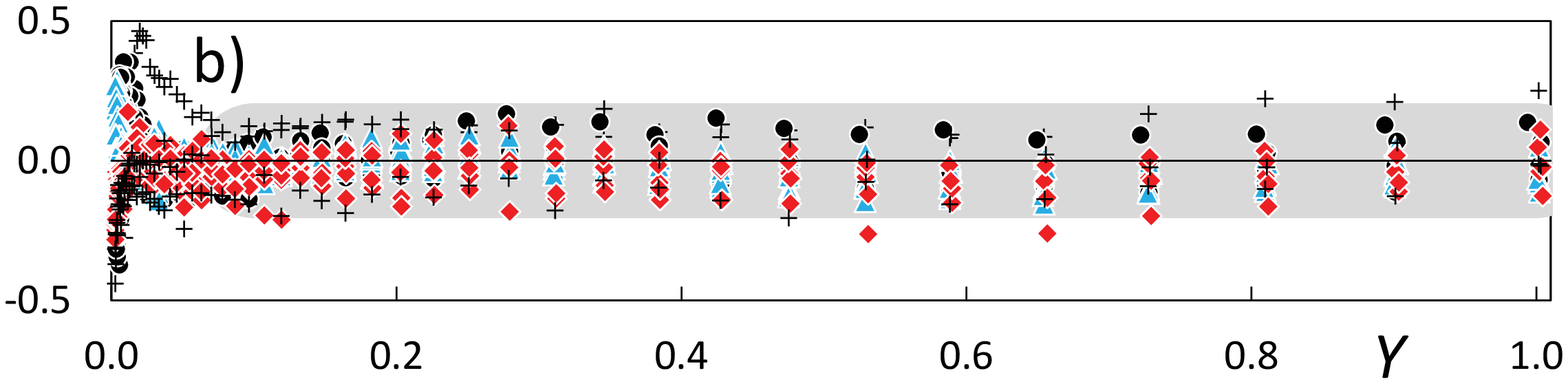}
\caption{(color online) (a) Superpipe profiles of figure \ref{Fig:Plog1} plus $\Delta U^+_{\mathrm{log}}$ (equ. \ref{UPlift}). Thin grey lines, $(U^+_{\mathrm{outer}} - U^+_{\mathrm{logint}} + \Delta U^+_{\mathrm{log}})$ equ. (\ref{Pout}).  $\cdot\cdot\cdot$, leading term $(5.68\, Y^2)$ of the Taylor expansion of $(U^+_{\mathrm{outer}} - U^+_{\mathrm{logint}} + \Delta U^+_{\mathrm{log}})$ for $R^+ = 3.31\times 10^3$ (first black profile), $1.09\times 10^4$ (first blue profile), $4.21\times 10^4$ (last blue profile) and $1.65\times 10^5$ (last red profile).
(b) $(U^+ - U^+_{\mathrm{outer}})$ given by equ. (\ref{Pout}). The grey band indicates deviations of up to $\pm 0.2$ from zero. See caption of figure \ref{Fig:Plog1} for other symbols. }
\label{Fig:Plog3}
\end{figure}

The remaining task is to find appropriate fits for the profiles of figure \ref{Fig:Plog1}. As the difference between interior and exterior log-laws is seen to be well described by $- \Delta U^+_{\mathrm{log}}$ (equ. \ref{UPlift}), it is useful to first ``lift`` the data of figure \ref{Fig:Plog1} by $\Delta U^+_{\mathrm{log}}$. The result is shown in figure \ref{Fig:Plog3}a .

As both $U^+_{\mathrm{logint}}$ and $\Delta U^+_{\mathrm{log}}$ have been designed to satisfy the symmetry condition on the centerline the remaining wake in figure \ref{Fig:Plog3} is easily fitted by
\begin{equation}
\label{Pout}
U^+_{\mathrm{outer}}(Y)  - \left[U^+_{\mathrm{logint}} - \Delta U^+_{\mathrm{log}}\right] = 2.30\, \sin^2\left(\frac{\pi}{2} Y\right)  \quad .
\end{equation}
As seen in figure \ref{Fig:Plog3}b, $U^+_{\mathrm{outer}}(Y)$ fits all the outer Superpipe profiles with an absolute error below $\pm 0.2\,$. From the small-$Y$ expansion of $U^+_{\mathrm{outer}}$ it follows that the initial deviation from the compound log-law $U^+_{\mathrm{logint}} - \Delta U^+_{\mathrm{log}}$ in the pipe is much more gradual than in the ZPG TBL, i.e.\, $\propto Y^2$ versus $\propto \eta^4$ in the ZPG TBL.

\section{The mean velocity profile in channels and ducts}
\label{sec:channel}

Below $\Reytau \equiv H^+ \approx 1000$ ($H^+$ being the non-dimensional channel/duct half height) the $U^+$-profiles in ducts become progressively affected by the duct aspect ratio, as documented by \citet{VinEF14}, \citet{VinJT14} and \citet{Vinuesa2016300}. Hence, attention is focussed here on the Reynolds number range $H^+ \geq 950$ where the difference between $U^+$ from channel DNS and duct experiments is of the order of the experimental uncertainty. For the following fits, six profiles from channel DNS's of \citet{HJ06}, \citet{LJ14}, \citet{LM14} and \citet{TMG14} with $950 \leq H^+ \leq 5200$ are used, together with four experimental profiles of \citet{ZDN03} for $2155 \leq H^+ \leq 4783$ and another four of \citet{SchultzFlack2013} for $1010 \leq H^+ \leq 5900$.

\begin{figure}
\center
\includegraphics[width=0.65\textwidth]{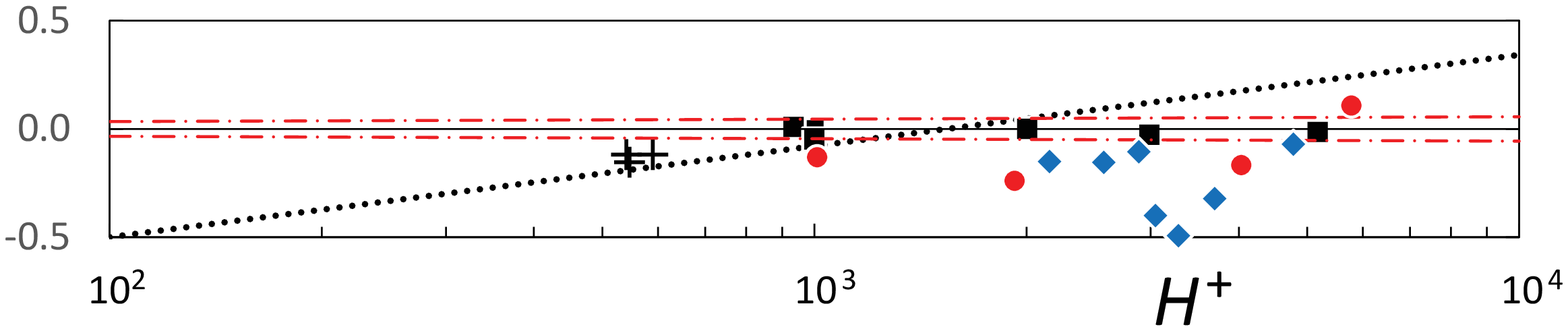}
\caption{(color online) Channel/duct centerline velocity of the profiles in fig. \ref{Fig:Clog} minus $U^+_{\mathrm{CL}}$ (equ. \ref{CCL}) versus $H^+$. $\blacksquare$, DNS channel data for $H^+ \geq 950$; $\bullet$ (red), \citet{SchultzFlack2013}; $\blacklozenge$ (blue), \citet{ZDN03}. See caption of fig. \ref{Fig:Clog} for more details. $+$, DNS data at $H^+ = 550$ and $590$ not considered for the fit (\ref{CCL}). $\cdot - \cdot$ (red), $\pm 0.2\%$ of $U^+_{\mathrm{CL}}$ (equ. (\ref{CCL}). $\cdot\cdot\cdot$, slope corresponding to
 $\kappa =0.384$.}
\label{Fig6}
\end{figure}

For channels and ducts there are no data at high enough $H^+$ to reveal the exterior log region as in figure \ref{Fig:Plog1} for the Superpipe. As a matter of fact, the available $H^+$ are still too low to even see a clear log-law in the $U^+(y^+)$ profiles and so there is not yet any controversy about a difference between $\kappa$'s extracted from the near-wall logarithmic region and from $U^+_{\mathrm{CL}}$. The fortunate difference to the pipe is the availability of high quality DNS channel data over the entire $H^+$-range of the laboratory experiments, which allow a reliable fit of $U^+_{\mathrm{CL}}$, despite the limited $H^+$-range. The best fit for the DNS channel data above $H^+=950$, with an $R^2$ value of 0.9998 is
\begin{equation}
U^+_{\mathrm{CL}} = \frac{1}{0.413}\, \ln(H^+) + 5.88
\label{CCL}
\end{equation}
As seen in figure \ref{Fig6}, the experimental $U^+_{\mathrm{CL}}$ have a much larger scatter and fall generally below the DNS fit, which is interpreted as a remnant of the aspect ratio effect.

\begin{figure}
\center
\includegraphics[width=0.65\textwidth]{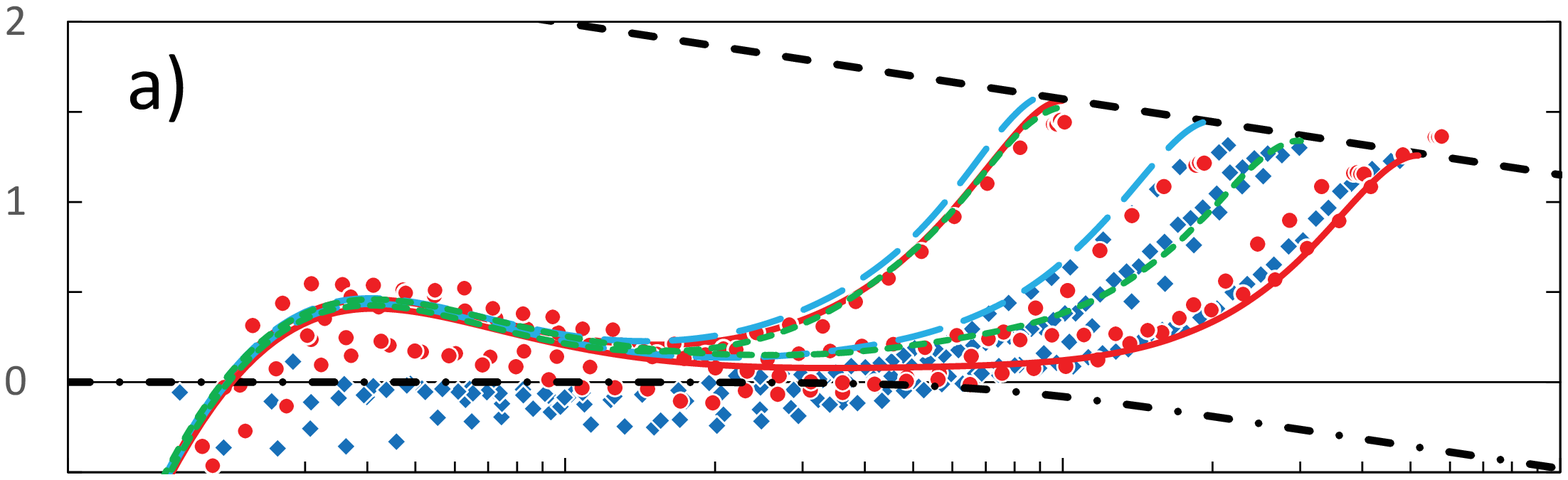}
\includegraphics[width=0.65\textwidth]{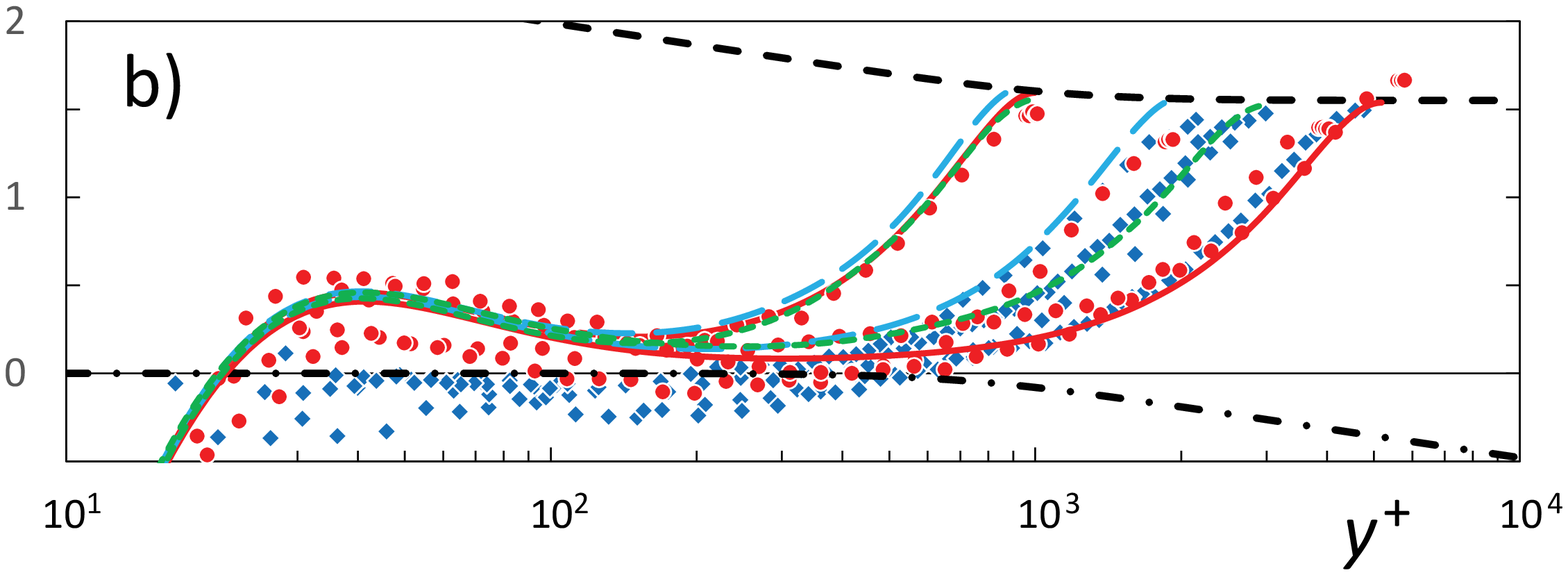}
\includegraphics[width=0.65\textwidth]{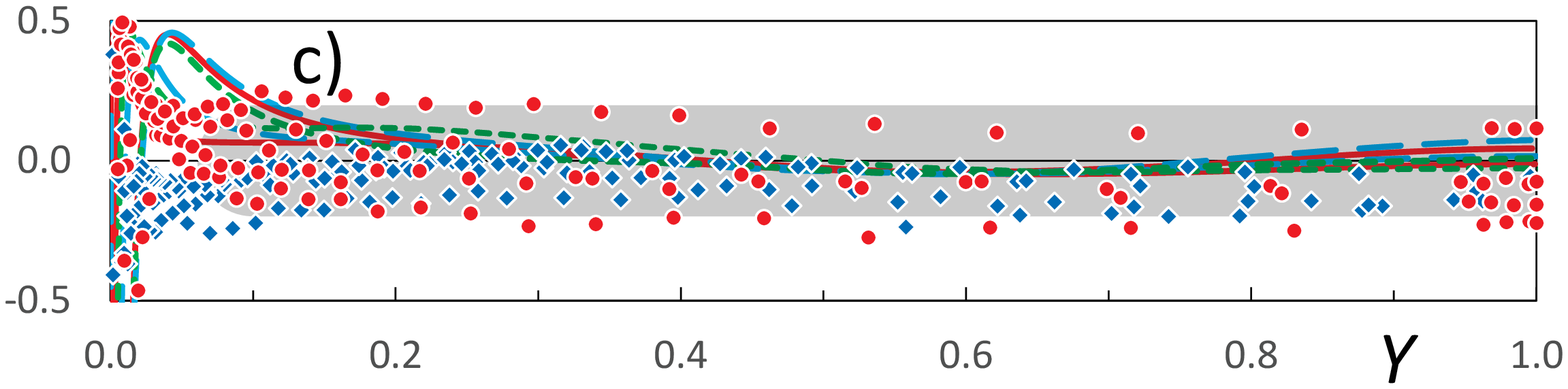}
\caption{(color online) Six DNS profiles of \citet{HJ06} (long blue dashes), \citet{TMG14} (short green dashes) and \citet{LM14} (red lines), four profiles of \citet{SchultzFlack2013} (red $\bullet$) and four profiles of \citet{ZDN03} (blue $\blacklozenge$).
(a) $(U^+ - U^+_{\mathrm{logint}})$ (equ. \ref{Plogint}) versus $y^+$; $-\cdot -$, $(-\Delta U^+_{\mathrm{log}})$ given by equ. (\ref{UClift}) for $H^+ \to \infty\,$;
\,$-\,-\,-$, Centerline velocity (\ref{CCL}) minus the  interior log-law (\ref{Plogint}).
(b) Data of fig. (a) plus $\Delta U^+_{\mathrm{log}}$ (equ. \ref{UClift}).
(c) $(U^+ - U^+_{\mathrm{outer}})$ given by equ. (\ref{Cout}). The grey band indicates deviations of up to $\pm 0.2$ from zero. }
\label{Fig:Clog}
\end{figure}

Focussing on the DNS by \citet{LM14} at $H^+=5200$, a nascent near-wall log-law with the ZPG TBL parameters is evident in figure \ref{Fig:Clog}a, but there is no evidence of an outer log-law with $\kappa = 0.413$ to match $U^+_{\mathrm{CL}}$. From the asymptotic matching argument in section \ref{sec:intro}, one must therefore conclude that, at the available $H^+$'s, the outer log-law is ``telescoped'' into the wake and the two regions will only separate at considerably higher $H^+$'s. This does however not preclude the use of the same asymptotics-inspired fitting scheme as for the pipe, since asymptotic expansions generally provide useful approximations already before the different regions (here the exterior log region and the wake) are clearly separated (see \eg the excellent outer fit of the low-$R^+$ data in fig. \ref{Fig:Plog3}a.

Hence, the universal interior log-law $U^+_{\mathrm{logint}}$ (equ. \ref{Plogint} with $R^+$ replaced by $H^+$) is first subtracted from the data and the remainder is ``lifted'' by the channel equivalent of equ. (\ref{UPlift})
\begin{eqnarray}
\label{UClift}
\Delta U^+_{\mathrm{log}} &=& \frac{1}{3}\left(\frac{1}{0.384} - \frac{1}{0.413}\right) \ln\left\{1 + \left[\frac{0.0014}{0.5\,\pi}\,H^+ \sin\left(\frac{\pi Y}{2}\right)\right]^3\right\} \\
&\sim & 0.183 \,\ln\left\{0.0014\,y^+\right\} - 0.0752 \,Y^2  + \mathcal{O}\left(Y^4\right) \quad \mbox{for} \quad \left(0.0014 \, H^+\right)^{-1} \ll Y \ll 1 \nonumber
\end{eqnarray}
Note that in equation (\ref{UClift}) the switch-over point between the wall log layer and the log layer with $\kappa = 0.413$ has been placed at $y^+ = (0.0014)^{-1} \approx 700$, as opposed to $\approx 500$ in the pipe (see discussion in section \ref{sec:conclusion}). The remainder is then fitted analogous to (\ref{Pout}):
\begin{equation}
\label{Cout}
U^+_{\mathrm{outer}}(Y)  - \left[U^+_{\mathrm{logint}} - \Delta U^+_{\mathrm{log}}\right] = 1.55\, \sin^2\left(\frac{\pi}{2} Y\right)  \quad .
\end{equation}

The lifted profiles are shown in figure \ref{Fig:Clog}b and the excellent quality of the complete outer fit $U^+_{\mathrm{outer}}$ for the DNS channel profiles can be appreciated in figure \ref{Fig:Clog}c. As in figure \ref{Fig6}, all but one experimental $(U^+ - U^+_{\mathrm{outer}})$ remain slightly negative at all $Y$, most probably due to finite duct aspect ratios.

\section{Discussion}
\label{sec:conclusion}

The starting point for the present study was the observation that it is hardly possible to fit the free-stream velocity in the ZPG TBL (equ. \ref{TBLout} with $\eta \to \infty$) and the centerline velocities in channels (fig. \ref{Fig6} and equ. \ref{CCL}) and pipes (fig. \ref{Fig:PCL} and equ. \ref{PCL}) with a single universal K\'arm\'an constant. From standard asymptotic matching principles (see section \ref{sec:intro}) it follows, that the $\kappa_{\mathrm{ext}}$ of the external log-laws matching up to the wake regions of the three canonical flows must also be different and related to the different pressure gradients.
Indeed, the parameters of the three exterior log layers $U^+_{\mathrm{logext}} = \kappa_{\mathrm{ext}}^{-1} \ln(y^+) + B_{\mathrm{ext}}$ extracted in this study, i.e. $\kappa_{\mathrm{ext}}$ = 0.384, 0.413 and 0.42 and the corresponding additive constants $B_{\mathrm{ext}}$ = 4.22, 5.42 and 5.60,  are found to be a monotonically increasing function of the favorable pressure gradient. Furthermore, they closely follow the correlation
\begin{equation}
\label{param}
\kappa_{\mathrm{ext}} B_{\mathrm{ext}} = 1.6\,[\exp(0.163\,B_{\mathrm{ext}})\,-\,1] \quad \mbox{and}
\end{equation}
which is virtually identical to equation (13) of \citet{NagibChauhan2008} based on a wide range of pressure gradient boundary layers. It is worth noting that the positive curvature of the graph $\kappa_{\mathrm{ext}} B_{\mathrm{ext}}$ versus $B_{\mathrm{ext}}$ necessarily implies the monotonic increase of $\kappa_{\mathrm{ext}}$ with increasing favorable pressure gradient, which is at odds with the low $\kappa$ given by \citet{NagibChauhan2008} for the channel.

Beyond the correlation (\ref{param}), the increase of $\kappa_{\mathrm{ext}}$ relative to its value in ZPG TBL's can now be directly related to the pressure gradient $(\dd p^+/\dd x^+)$, with $p^+ \equiv \widehat{p}/(\widehat{\rho}\,\widehat{u}_\tau^2)$, which is equal to $1/\Reytau$ and $2/\Reytau$ for channel and pipe flow, respectively. The simple fit
\begin{equation}
\label{kappa}
\kappa - \kappa_{\mathrm{ZPG}} = 0.029\,\left(-\Reytau \frac{\dd p^+}{\dd x^+}\right)^{1/3}\quad ,
\end{equation}
is good, but needs to be taken ``with a grain of salt'' as it is based on only two points and the uncertainty of $\kappa - \kappa_{\mathrm{ZPG}}$ is large.

So, what remains of universality is the buffer layer and an interior log-law with the same (within fitting accuracy) $\kappa_{\mathrm{int}} = 0.384$ and additive constant $B_{\mathrm{int}} = 4.22$ as in the ZPG TBL. However, this region with universal $U^+(y^+)$ has been found to shrink with increasing pressure gradient : while it extends to $y^+_{\mathrm{int-ext}} \approx 700$ in the channel, its extent is reduced to $\approx 500$ in the pipe. A summary of the parameters of the interior and exterior log-laws, together with the additive constants $C$ in the log-laws for $U^+_\infty$ and $U^+_{\mathrm{CL}}$ are collected in table \ref{Table1}.

\begin{table}
\center
\begin{tabular}{l c c c c c c c c}
 & $-\Reytau (\dd p^+/\dd x^+)$ & $\kappa_{\mathrm{int}}$ & $B_{\mathrm{int}}$ & $y^+_{\mathrm{int-ext}}$ & $\kappa_{\mathrm{ext}}$ & $B_{\mathrm{ext}}$ & $C$ \\
 ZPG TBL & 0 & 0.384 & 4.22 & $-$ & 0.384 & 4.22 & 3.26 \\
 Channel & 1 & 0.384 & 4.22 & $\approx$ 700 & 0.413 & 5.42 & 5.88 \\
 Pipe & 2 & 0.384 & 4.22 & $\approx$ 500 & 0.420 & 5.60 & 6.84 \\
\end{tabular}
\caption{Summary for the three canonical flows of the parameters for the interior log-law $\kappa_{\mathrm{int}}^{-1}\ln(y^+) + B_{\mathrm{int}}$, the switch-over point $y^+_{\mathrm{int-ext}}$ from interior to exterior log-law, the parameters for the exterior log-law $\kappa_{\mathrm{ext}}^{-1}\ln(y^+) + B_{\mathrm{ext}}$ and the additive constant for $U^+_\infty$, respectively $U^+_{\mathrm{CL}} = \kappa_{\mathrm{ext}}^{-1}\ln(\Reytau) + C$.}
\label{Table1}
\end{table}

The proposed double log-layer structure for wall-bounded flows with weak pressure gradients, specifically pipe and channel flows, resolves the apparent conflict between the log-law parameters extracted from centerline data and those obtained from near-wall velocity profiles, especially from profiles at Reynolds numbers below $\Reytau \approx 10^4$, where in the overwhelming majority of experiments and computations the K\'arm\'an constant $\kappa$ is found to be within $\pm 0.005$ of the ZPG TBL value of $0.384$.

As mentioned before, one can qualitatively understand this double log-layer structure from the stream-wise momentum balance, in which the importance of the pressure gradient increases with wall distance. However, it is not clear how the switch-over point $y^+_{\mathrm{int-ext}}$ between interior and exterior log-laws should scale. Based on the available data it cannot scale on the intermediate variable $y^+/\Reytau^{1/2}$ and, therefore, flow specific fixed values of $y^+_{\mathrm{int-ext}}$ have been used in the present fits (\ref{UPlift}) and (\ref{UClift}). Note also that in the channel the exterior log-law remains ``buried'' in the wake at the available $H^+$, so that $y^+_{\mathrm{int-ext}} \approx 700$ can only be a rough estimate. Despite the large uncertainty of these switch-over points, the trend appears clear and consistent with an increasing pressure gradient becoming significant in the momentum balance closer and closer to the wall.

Considering the observed trend of $y^+_{\mathrm{int-ext}}$ with pressure gradient, one may speculate that the interior logarithmic region will disappear for pressure gradients not much larger than in the pipe. Conversely, one would expect an expansion of the universal interior logarithmic region when reducing the pressure gradient below the channel value. Exploring these ideas experimentally appears exceedingly difficult, but it may be interesting to explore high Reynolds number channel flows with artificially altered friction on one of the walls by high resolution DNS.

\begin{acknowledgements}
I am most grateful to Hassan Nagib for his continuous encouragement and even more so for initiating this research by relating to me his discussion with Don Coles, in which Don asked him why he was bothering to extract $\kappa's$ from velocity profiles when it was so much easier and more reliable to do it from free-stream or centerline data. I also thank Hassan Nagib, Bob Moser, MyoungKyu Lee and Mike Schultz for sharing data and Ivan Marusic for communicating the data file corresponding to figure 1 in \cite{MMHS13}.
\end{acknowledgements}

\appendix
\section{Alternative fit of the Superpipe data with a single log-law and $\kappa = 0.384$}
\label{sec:app}

As discussed in sections \ref{sec:intro} and \ref{sec:pipe}, the asymptotic matching argument for equal $\kappa$'s in the expression (\ref{PCL}) for the pipe centerline velocity and in the log-law adjacent to the wake hinges on the assumption that there are no additional measurable Reynolds-number-dependent terms in (\ref{PCL}). This assumption is strongly supported by the following failed attempt to develop a Reynolds number dependent fit for the difference between $U^+$ and $U^+_{\mathrm{logint}}$ with $\kappa = 0.384$, given by equation (\ref{Plogint}).

\begin{figure}
\center
\includegraphics[width=0.65\textwidth]{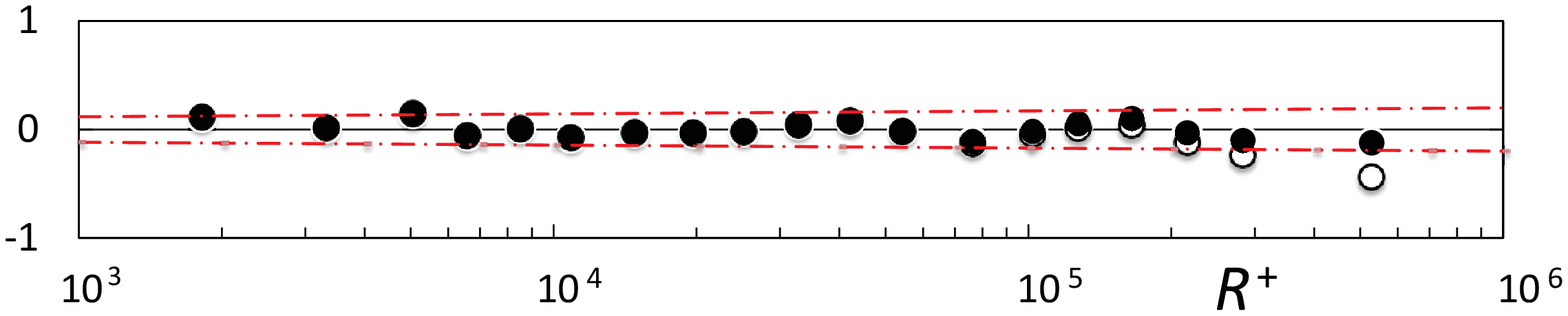}
\caption{(color online) Superpipe Pitot centerline velocities minus $U^+_{\mathrm{CL,alt}}$ (equ. \ref{ACL}) versus $R^+$. $\bullet$, $\circ$, same as in figure \ref{Fig:PCL}. $\cdot - \cdot$ (red), $\pm 0.5\%$ of $U^+_{\mathrm{CL,alt}}$.}
\label{Fig:ACL}
\end{figure}

Starting again with the centerline, the fit
\begin{equation}
U^+_{\mathrm{CL,alt}} = \frac{1}{0.384}\, \ln(R^+) + 2.35 + \frac{22}{\ln(R^+)} \quad ,
\label{ACL}
\end{equation}
is seen in figure \ref{Fig:ACL} to be as good as (\ref{PCL}).

\begin{figure}
\center
\includegraphics[width=0.65\textwidth]{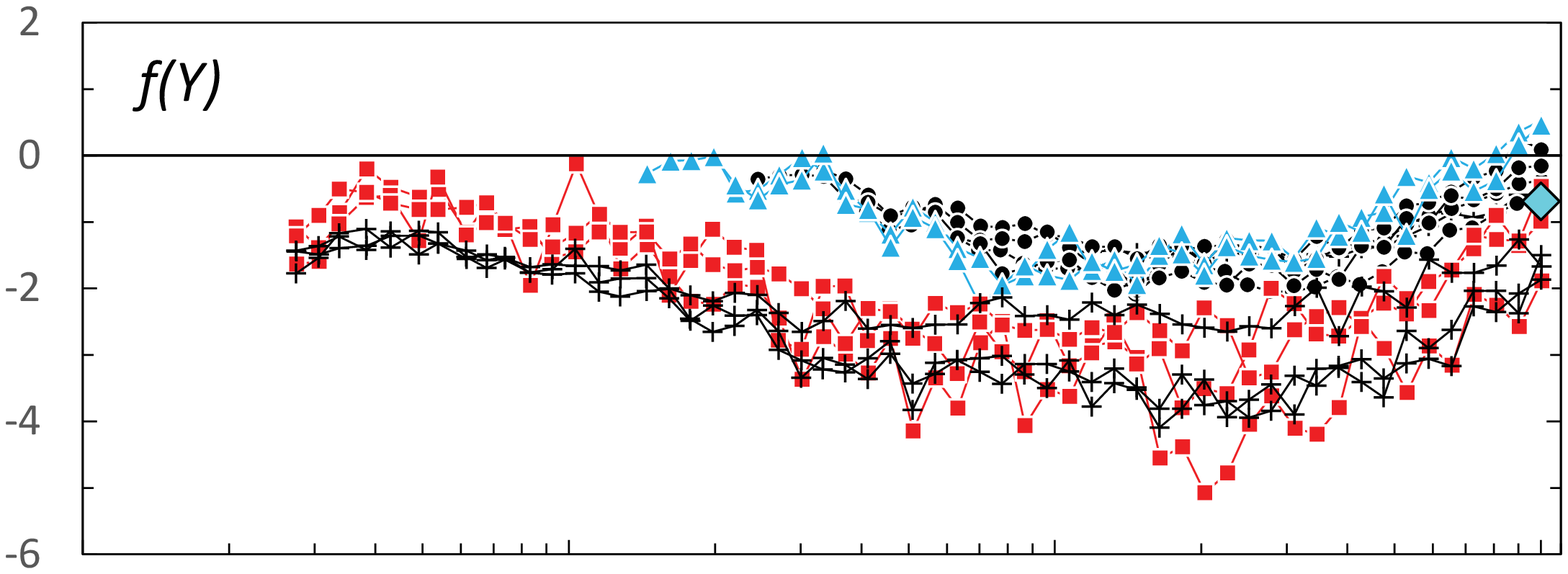}
\includegraphics[width=0.65\textwidth]{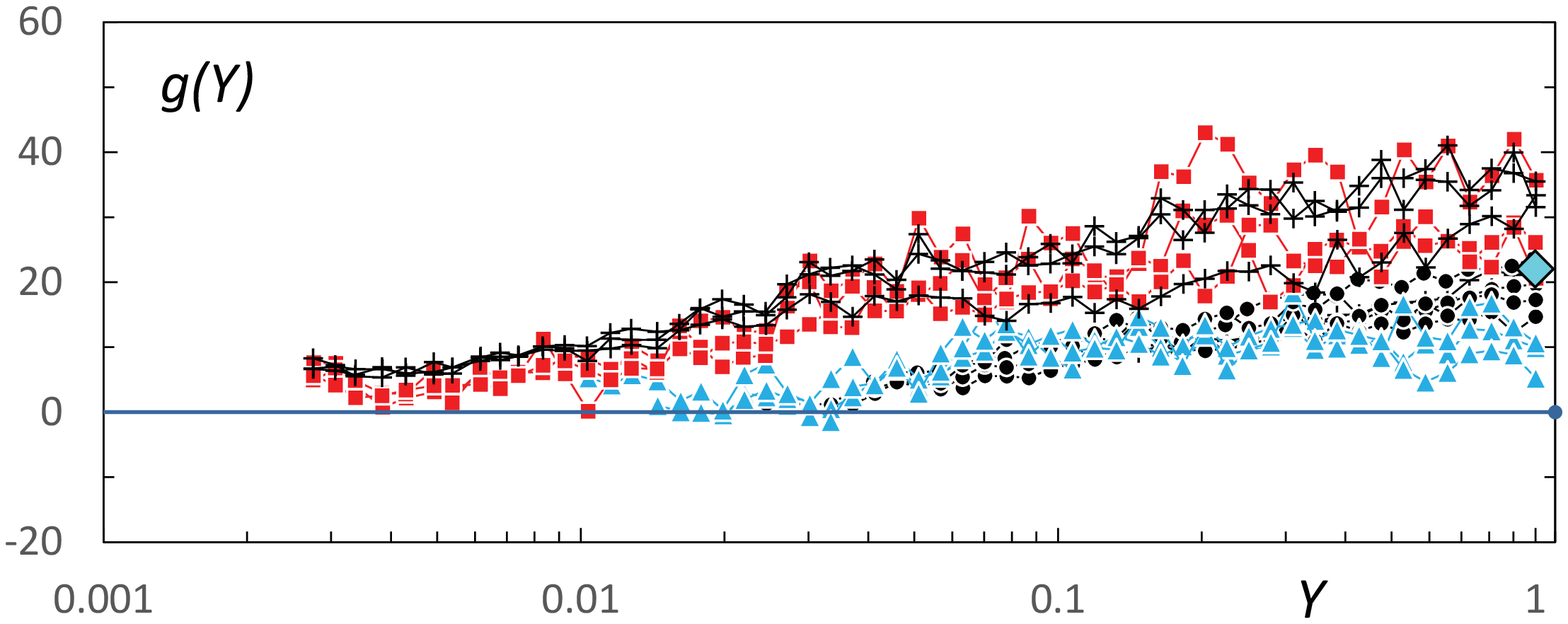}
\caption{(color online) $f(Y)$ and $g(Y)$ of the alternate outer fit \ref{Awakefit} obtained by pairing the seven lowest and the six highest $R^+$ of fig \ref{Fig:Plog1} with the reference 11th profile ($R^+ = 4.21\times 10^4$).
$\blacklozenge$ (aqua), $f(1)=-0.69$ and $g(1)=22$ required to match the centerline fit \ref{ACL}. Other symbols as in fig. \ref{Fig:Plog1}.}
\label{Fig:Afg}
\end{figure}

A possible outer fit reducing to (\ref{ACL}) on the centerline is
\begin{equation}
U^+ = \frac{1}{0.384}\, \ln \left\{\frac{5.05}{0.5\,\pi}\,R^+\,\sin \left(\frac{\pi\,Y}{2}\right)\right\} + f(Y) + \frac{g(Y)}{\ln(y^+)}
\label{Awakefit}
\end{equation}
implying that $f(1)=-0.69$ and $g(1)=22$. These two functions can be determined from pairs of profiles $U^+ - U^+_{\mathrm{logint}}$ (figure \ref{Fig:Plog1}). Pairing the seven lowest and the six highest $R^+$ with the 11th profile for $R^+ = 4.21\times 10^4$ used as reference, the resulting $f$ and $g$ for each pair are shown in figure \ref{Fig:Afg}.

It is obvious that the procedure to extract $f$ and $g$ from the experimental profiles produces very noisy results in figure \ref{Fig:Afg}, but it does nevertheless reveal a clear Reynolds number trend not compatible with the assumed fit (\ref{Awakefit}). The difference between the $f(Y)$ obtained from low and high Reynolds number profiles is particularly marked in the region $y \lesssim 0.1$, i.e. in the region of the exterior log-law with $\kappa =0.42$ of figure \ref{Fig:Plog1}, where the difference between $f(Y)$ from low and high Reynolds number pairs reaches 2, whereas the new fit (\ref{Pout}) built on two log-laws fits the data within $\pm 0.2$, as seen in figure \ref{Fig:Plog3}. While this failed attempt can obviously not exclude the possibility of finding a better Reynolds dependent wake function for the difference between $U^+$ and a single universal log-law with $\kappa \simeq 0.38 - 0.39$, it renders this option rather unlikely.

\bibliographystyle{jfm}
\bibliography{Karmanbib}
\end{document}